# Automatic Handover Control for Distributed Load Balancing in Mobile Communication Network

Dina Farouk ALtayeb Ali[1], Fatima abdelgani Mustafa[1,] Mohamed Osman Mustafa Fdolelsid[1]

[1]Future University, Faculty of Engineering, Department of Computer Engineering

**ABSTRACT:**

Due to technology and its daily continuous advancing and enhancement the users dealing with it are hungry for the best and better quality of service and this demand never stops, from this point came the idea of my research which aims to enhance the mobile network using the LOAD BALANCE and its algorithm to make it better ,in my research am requiring a HANDOVER process to be involved to get the best results , the simulation was implemented by MATLAP software to implement the two systems the normal system and the system with the load balance, result and graphs also came from the MATLAP software, one of the load balance methods was used and it's the round robin method with all the needed variables based on the quantum or time slice. It's the round robin method that helped to get better results first reducing the execution timing by (722626 MS) and improvement in the blocking probability and reduce it by 20%

**Introduction:**

Due to rapid change in technology the demand for better and faster cellular communication also increases. This growth in field of cellular communication has led to increase intensive research toward and development toward cellular system. The main reason of this growth is newly concept of mobile terminal and user mobility. The main characteristics of cellular communication system offer user maximum freedom of movement while using cell phones (mobiles). A cellular network is made up of numbers of cells (or radio cells). Each cell is allocated a band of frequencies and served by base station consisting of transmitter, receiver and control unit. Adjacent cells are assigned different frequencies to avoid interference or cross talk [1]. As more customers use the cellular network with single base station traffic may be build up so there are not enough frequency bands assigned to a cell to handle its calls [2].

Load balancing is a computer networking method for distributing workloads across multiple computing resources, such as computers, a computer cluster, network links, central processing units or disk drives [3].

Load balancing aims to optimize resource use, maximize throughput, minimize response time, and avoid overload of any one of the resources. Using multiple components with load balancing instead of a single component may increase reliability through redundancy. Load balancing is usually provided by dedicated software or hardware, such as a multilayer switch or a Domain Name System server process [4].

The characteristics of mobile/wireless computing (i.e., limited bandwidth, periodic disconnection, battery constraints, etc.) have led to the development of a number of systems for providing improved network access in such environments. With the rapid increase in the use of such systems, however, new problems have now arisen. One such problem is that of overloading the base stations or "fixed units" (FUs) where mobile users connect to the network with the computational work of performing optimizations such as encoding and encryption. This problem is particularly true of systems that provide aggressive optimization of wireless transmissions such as the one proposed by Walky talky and later refined by Graham et al.

At peak times, in dense urban areas, many mobile units (MUs) may be connected to the same FU and the load induced by the necessary transmission optimizations may exceed the FU's capabilities. When this happens, any potential benefits of the optimizations may be effectively negated [5].

Mobility is the most important feature of a wireless cellular communication system. Usually, continuous service is achieved by supporting handover from one cell to another. In cellular telecommunications Handover is the process of changing the channel (frequency, time slot, spreading code, or combination of them) associated with the current connection while a call is in progress. It is often initiated either by crossing a cell boundary or by a deterioration in quality of the signal in the current channel, in satellite communications it is the process of transferring satellite control responsibility from one earth station to another without loss or interruption of service. Handover is divided into two broad categories (hard and soft handoffs). They are also characterized by "break before make" and "make before break." In hard handovers, current resources are released before new resources are used; in soft handovers, both existing and new resources are used during the handover process. Poorly designed handover schemes tend to generate very heavy signaling traffic and, thereby, a dramatic decrease in quality of service (QoS). The reason why handovers are critical in cellular communication systems is that neighboring cells are always using a disjoint subset of frequency bands, so negotiations must take place between the mobile station (MS), the current serving base station (BS), and the next potential BS. Other related issues, such as decision making and priority strategies during overloading, might influence the overall performance [6].

There may be different reasons why a handover might be conducted:

- When the phone is moving away from the area covered by one cell and entering the area covered by another cell the call is transferred to the second cell in order to avoid call termination when the phone gets outside the range of the first cell;
- When the capacity for connecting new calls of a given cell is used up and an existing or new call from a phone, which is located in an area overlapped by another cell, is transferred to that cell in order to free-up some capacity in the first cell for other users, who can only be connected to that cell;
- In non-CDMA networks when the channel used by the phone becomes interfered by another phone using the same channel in a different cell, the call is transferred to a different channel in the same cell or to a different channel in another cell in order to avoid the interference;
- Again in non-CDMA networks when the user behavior changes, e.g. when a fast-travelling user, connected to a large, umbrella-type of cell, stops then the call may be transferred to a smaller macro cell or even to a micro cell in order to free capacity on the umbrella cell for other fast-traveling users and to reduce the potential interference to other cells or users (this works in reverse too, when a user is detected to be moving faster than a certain threshold, the call can be transferred to a larger umbrella-type of cell in order to minimize the frequency of the handovers due to this movement);

- in CDMA networks a handover (see further down) may be induced in order to reduce the interference to a smaller neighboring cell due to the "near-far" effect even when the phone still has an excellent connection to its current cell [7].

So what is the effect of the signal break and the hard handover process in the packet loss or time delay is the main problem of this theses.

**Mobile communication:**

It is communication network (either public or private) which doesn't depend on any physical connection between two communication entities and have flexibility to be mobile during communication. The current GSM and CDMA technologies offer Mobile Communication [10].

**Global system for mobile (GSM):**

GSM is a second generation digital cellular system. Digital transmission was used rather than analog transmission in order to improve transmission quality, system capacity, and coverage area. GSM works on three frequencies 900 MHz, 1800 MHz and 1900 MHz To make efficient use of frequency bands GSM networks uses combination of FDMA (frequency division multiple access) and TDMA (time division multiple access).GSM operators have set up roaming agreement with foreign operator which help users to travel abroad and use their cellphones Since many GSM network operators have roaming agreements with foreign operators, users can often continue to use their mobile phones when they travel to other countries. SIM cards (Subscriber Identity Module) holding home network access configurations may be switched to those with metered local access, significantly reducing roaming costs while experiencing no reductions in service.

GSM, together with other technologies, is part of the evolution of wireless mobile telecommunications that includes High-Speed Circuit-Switched Data (HCSD), General Packet Radio System (GPRS), Enhanced Data GSM Environment (EDGE), and Universal Mobile Telecommunications Service (UMTS) [18].

**GSM Network structure:**

The GSM system consist of several functional elements including mobile switching centers (MSC), base stations (BSC) with associated base transceivers (BTS), an operation and maintenance center (OMC) and gateway MSC.GSM mobile terminal or mobile stations communicates across the Interface, known as the air interface, with a base BTS in the small cell in which the mobile unit is located. This communication with a BTS takes place through the radio channels. The network coverage area is divided into small regions called cells. Multiple cells are grouped together form allocations area (LA) for the mobility management.

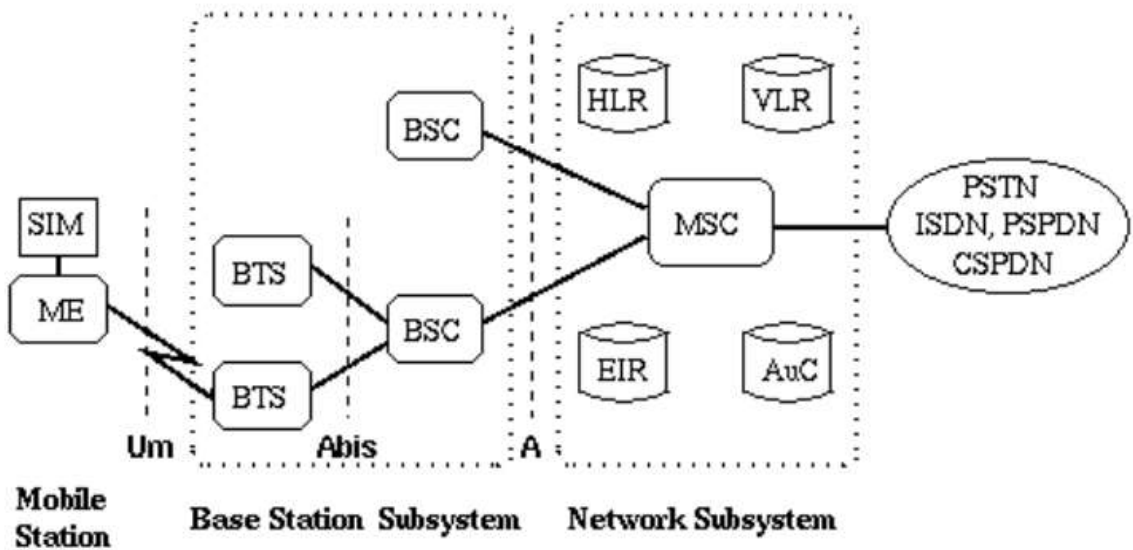

Figure 1: GSM structure

**Handover:**

It is the Process of transferring a call or data transfer in progress from one channel to another, the core network performs handovers at various levels of the system architecture or may handover the call to another network.

If the mobile device moves out of the range of one cell (base station) and a different base station can provide it with stronger signal and if all channels of one base station are busy then a nearby base station can provide service to the device [22].

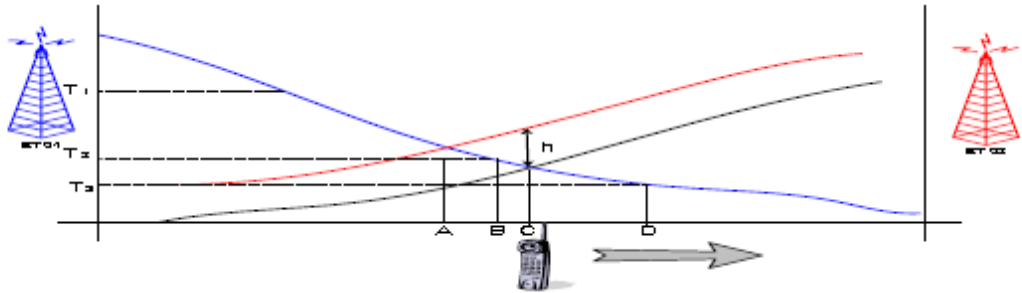

Figure 2: Handover scenario

**Handover techniques:**

- **Vertical Handover:** In mobile communications, vertical handover generally refers to the ability of a network operator to provide continuous service across different radio access networks or RATs. It means that the mobile user is unaware of the transfer of services during the procedure. Vertical handover thus provides a user with the ability to use services irrespective of the access network that is currently being used or that may soon be used. In technical terms, vertical handover refers to a user terminal changing the type of radio access technology it uses to access a supporting infrastructure, in order to support mobility. For example, a handset capable of both UMTS and GSM technologies will be able to handover between GSM and UMTS networks whenever necessary [22].

- **Horizontal handover:** GSM system use horizontal handover upon different scenarios listed as follow:

  I. **Intra BTS handover:** This occurs within the same BTS when there are some interference takes place. In this case mobile will be locked to the same BTS but the channel allocated to that mobile/time slot will change [23].

  II. **Inter-BTS Intra-BSC handover:** This type of handover occurs when the mobile moves out of the coverage of one BTS into another BTS and both BTSs are controlled by the same BSC. BSC will take care of the handover by allocating a channel for the user in the second BTS [23].

III. **Inter-BSC handover:** This is a special case of previous one and this time handover occurs between two BSCs. Therefore, it has to be controlled by MSC [23].

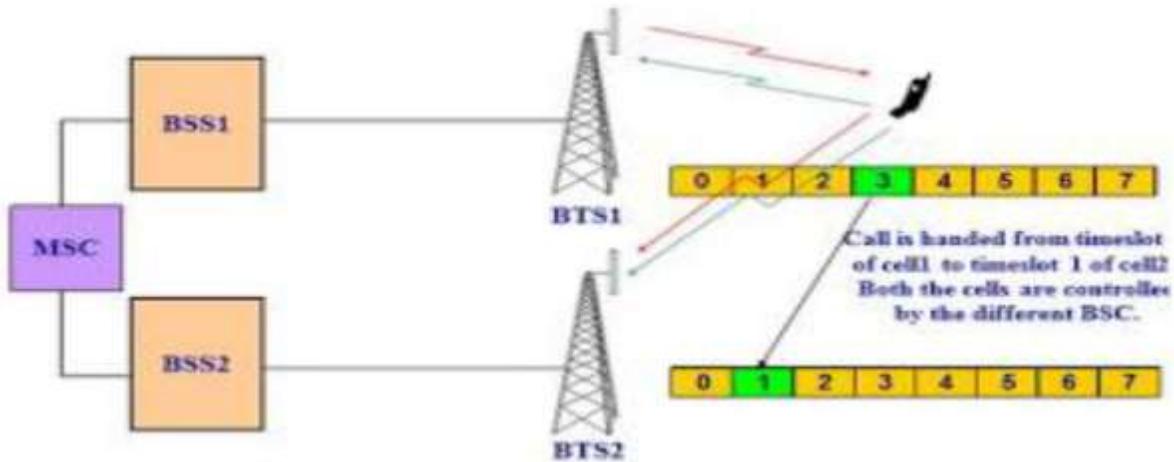

Figure 3: inter BSC handover

**Implementation:**

In the new system as shown in the block diagram the system is waiting for the calls request and when it comes it check the calls status if it's not overloaded the call will be assigned to the available channel to complete the call process.

if its overloaded the load balance process will take over by initialize a queue and determine the slice or quantum time for it and then checking the neighbor cell for availability and the inquire a handover process to happen and execute it for calling request depending on queue and until the quantum time finish then the system check if the process is over, If it's not over rearrangement of the unfinished process to the end of the queue until it's all finished, Then the system will make three checks the first one is to see if the queue is empty or not if not it will go back to make the necessary handover process and finish them, second check if the first one is clear it will check if all the calls requests are completed if yes the system will go back to wait for new requests if no that means that there are some calls requests the third check will take place to check if there is available channels for the calls if yes that will take it to initialize the queue ,if there is no available channels then the rest of the calls will be blocked and the system will be back to wait for new call requests.

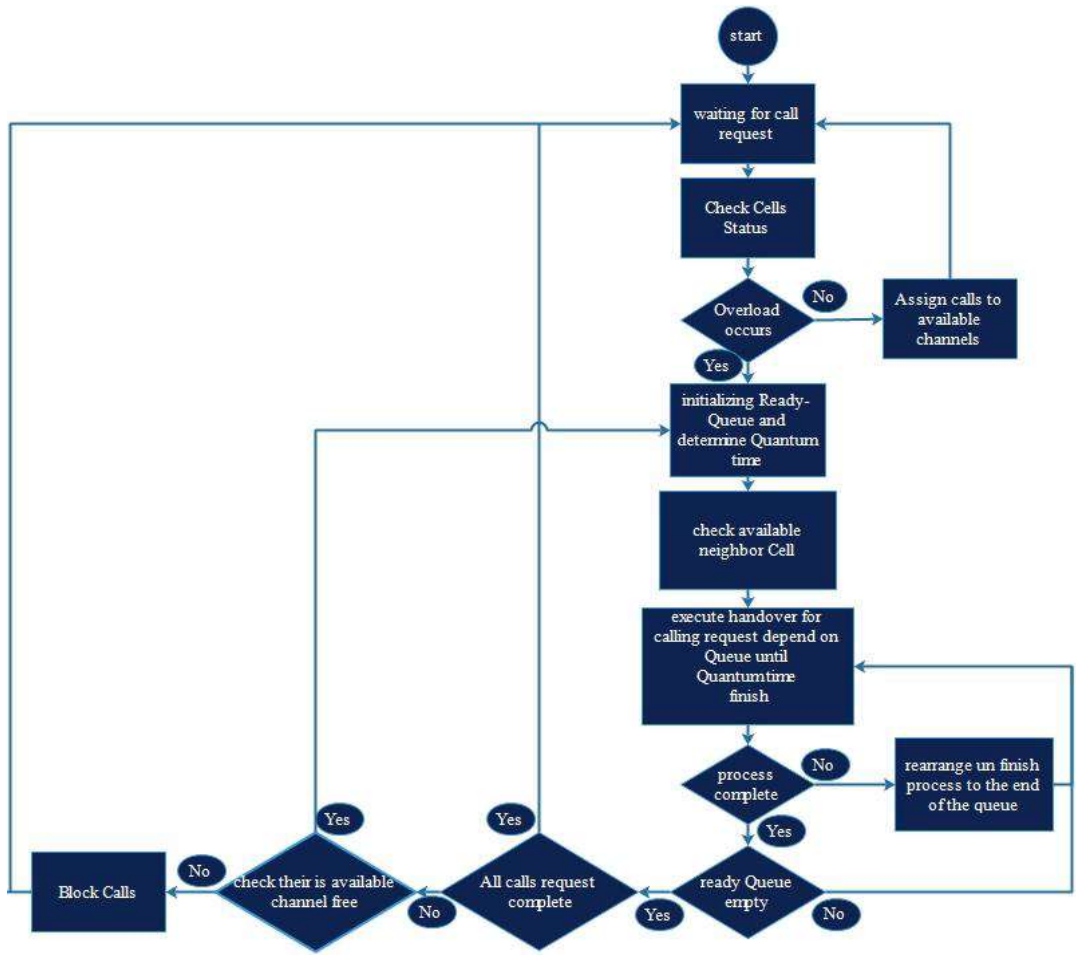

**Figure 4: Overall System Flow chart**

**Result:**

Computer simulation tests have been carried out to show the classifiers operation. All tests are done using a PC, P4B cor i5-2450M processor Ram 8GB. The code was developed under MATLAB2009 software, two simulation systems was designed
1. A system without load balancing (normal system)
2. A system with load balancing algorithm.

The simulation model of the cellular network area consists of one MSC, three BSC, and seven cells for each BSC. Using matlab to simulate the mobile network, the users distributed randomly in single BSC area of 1Km as in figure 5.

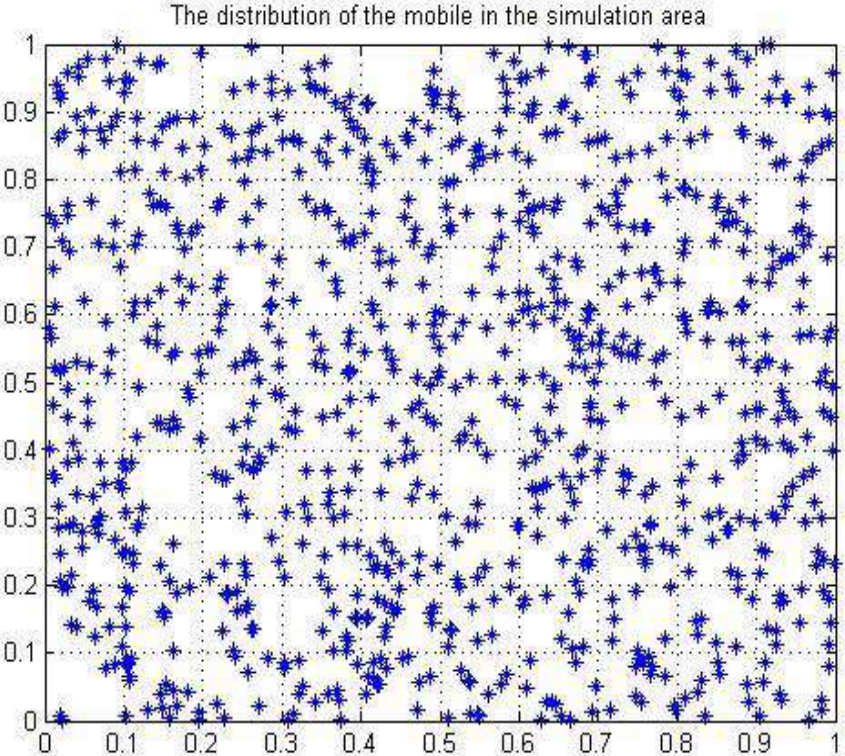

Figure 5: distribution of mobile in simulation area

This simulation is running on a major input parameter: Number of calling request where the user is asked to enter any number of his choose as shown in figure 6

```
Command Window
system have 7 cell per BSC

channel free BSC1 = 313

enter number of call request

900

 BSC1 overloaded
```

Figure 6: simulation Input parameter

**Execution time (delay):**

When a new call request is initiated either it will be accepted (forwarded to its destination), holed (waiting for acceptances) or rejected (when traffic over load occurs). The normal system work properly until the overload occur as simulated previously in figure 7, when overload occurred to the system without load balancing (Normal System) will block any calls that exceed the number of available channel in BSC1 (default) as shown in figure 7.

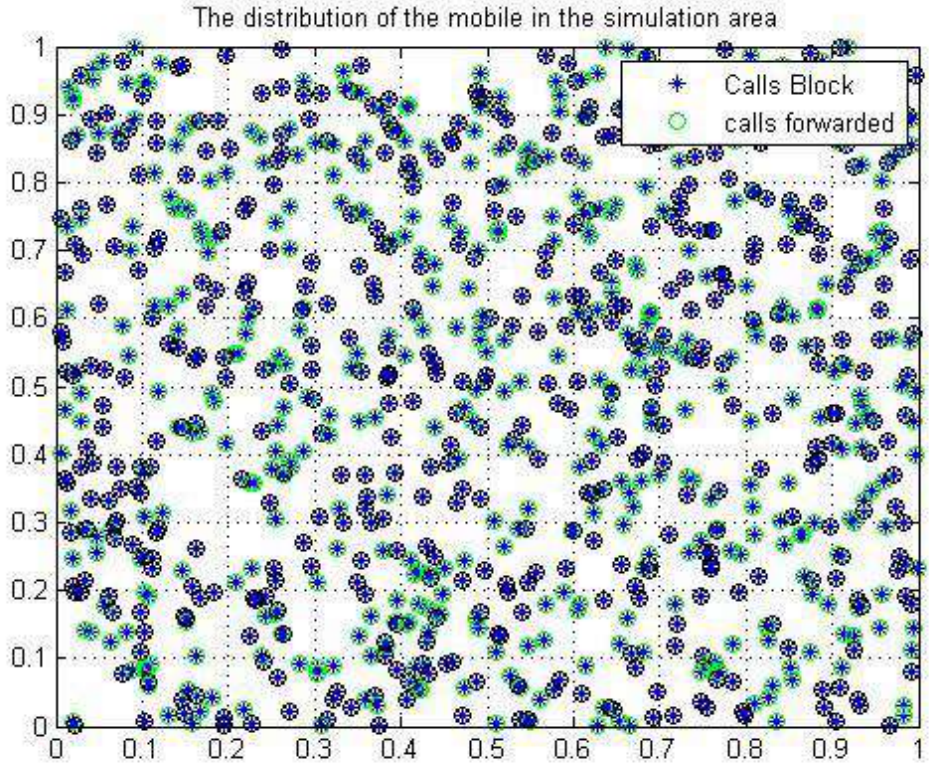

**Figure 6: Blocking calls in system without load balancing**

The performance of normal system will be enhanced by implementing load balancing algorithms, to handover overload calls to nearest cell in different BSC under same controller MSC.

The load balancing algorithm used in simulation is Round robin, which is working by initializing size of ready queue, the size of ready queue depends on the number of available channels in neighbor cells (BSC2, and BSC3), then set optimum quantum time $Q_{time}$ from formula:

$$Q_{time} = \frac{average\,arraival\,time\,range}{ready\,queue\,size}$$

$$\text{Quantum Time} = \frac{251.1959}{517.9749} = 0.4850 \text{ MS}$$

Round robin algorithm calculates the execution time for each call by adding quantum time (0.4850 MS) to context switch time (0.1MS), to get the actual time needed for each call to handover.

Execution time = Quantum time + context switch time

But normal system execution time calculate the execution time for each call by adding arrival time to waiting time 3 MS.

Execution time = arrival time + waiting time

The next figure 8 show the distribution handover calls between two neighbor BSC2 and BSC3 where the two cells are balanced in handling the handover calls without exceeding their maximum available channels.

```
Number of Handover calls  = 587

channel free BSC2 = 346

channel free BSC3 = 382

BSC2 Handeled = 296

BSC3 Handeled = 274
```

**Figure 8: handover calling distributed of mobile in neighbor**

figure 9 show the reducing blocking calls in load balancing system by distributed the overload among nearest available cell (BSC2 and BSC3).

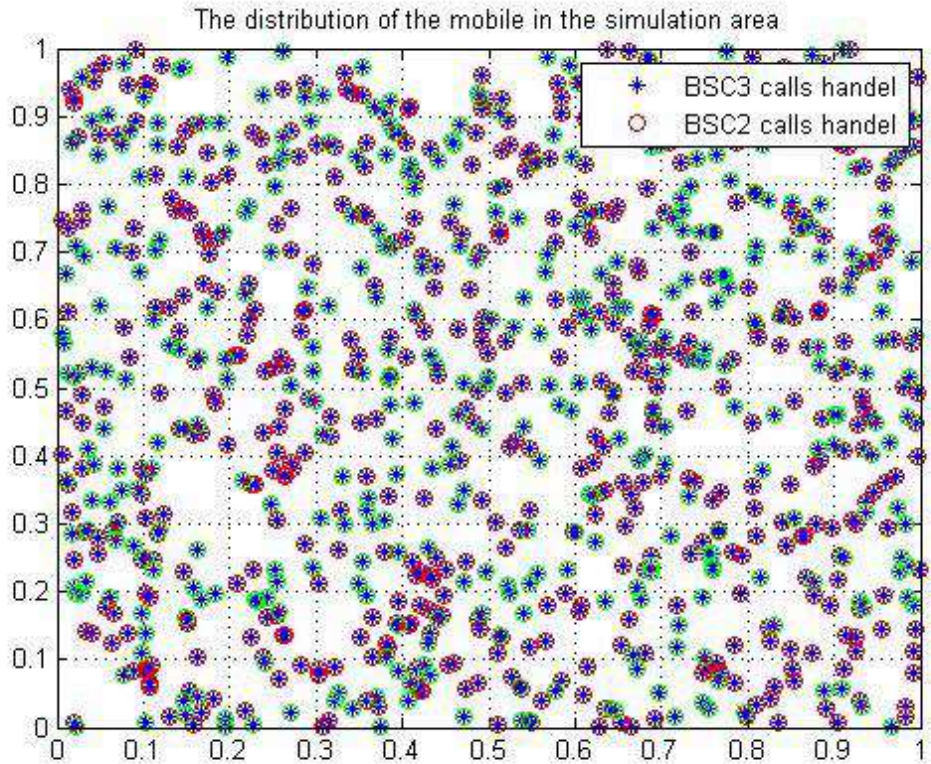

**Figure 9: handover calling distributed of mobile in simulation area**

Several metrics can be used to evaluate and compare the performance of the proposed algorithm. The call blocking probability is defined as the ratio of the number of new calls initiated by a mobile host which cannot be supported by existing channel arrangement to the total number of new calls initiated. Call blocking probability ($P_b$) is given by the ratio of "number of calls lost by the system" to "the total number of new calls initiated" [60].

The number of calls blocking rate for proposed algorithm is less than the conventional one because the load is distributed from over loaded cell to them neighbors light loaded cells so many of physical resource block will be free and able to accommodate new calls. To calculate blocking calls in each system using erlangb formula:

$$P_b = \frac{\frac{A^N}{N!}}{\sum_{i=0}^{N} \frac{A^i}{i!}}$$

Where: $P_b$=blocking call probability=Number of trunks in full availability group, A=Traffic offered to group in Erlangs:

Where A is given by $A = \lambda / \mu$, which is successive call time arrivals. In which, $\lambda$ is the call arrival rate per second, $\mu$ is average call departure rate of users per second, and N is number of channels in the system.

Calculating the blocking probability in order to compute the two system (NS and LBS) a different numbers of call requested are considered, Figure 10 show the relation between the calls blocking rate for two systems and the call request.

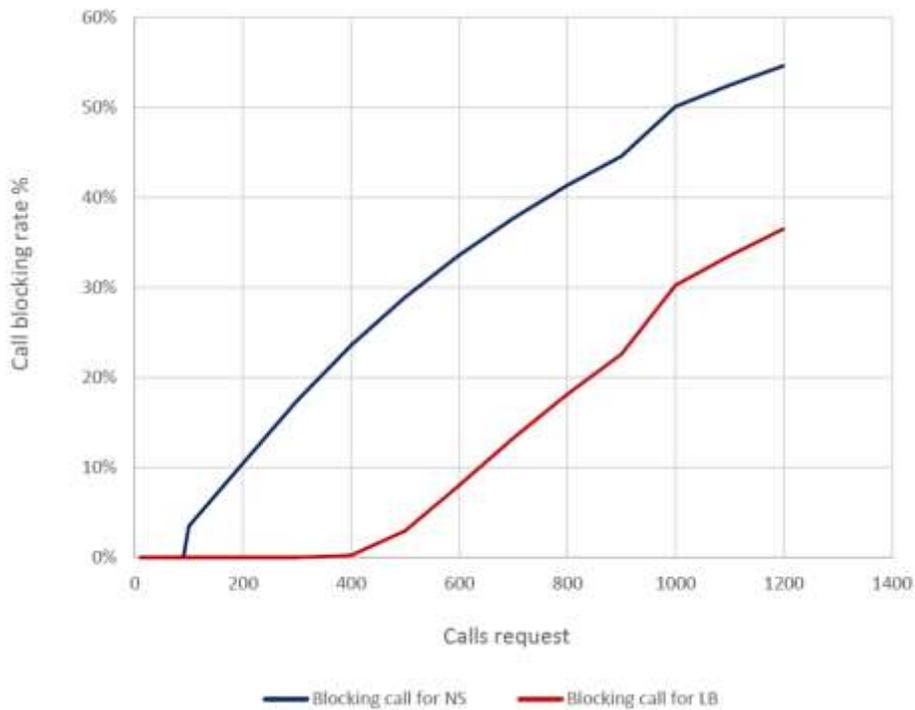

**Conclusion:**

This study of implementing Normal System of a GSM that suffers from a 2 major problems ( Delay and a Call blocking ) to be analysis and enchased in term of delay and calls by applying a load balance algorithm into a GSM system that showed enhancement of firstly: The performance that was measured in terms of delay (execution time), where the total timing measured for the Load Balance system (157931 MS) and Normal System (880557 MS) which approved that the applied Load Balance system has a less execution timing results that will minimize the holding time the call request waits for acceptance (forwarded).

secondly: Calculating the blocking probability in order to comparison the two systems (Norma System and Load Balance System) a different numbers of call requests in a range of (0-1200 call) are considered and their resulted blocking probability when it reached the maximum (1200) in the Normal System had a blocking probability of 55% Where the Load Balance system showed a less blocking probability of 37%.

**Recommendations and Future directive:**

The integrated LB system algorithm (RR) is applied regardless other parameters (signal strength, co-channel interference,…etc.) and others that could be taken in account in choosing other LB algorithm for farther future testing that also can be by inventing a new monitoring system of the MSC that prevent the overload to occur by applying an enhanced LBA that will have a great improvement and will follow the same principle procedures as proposed in this thesis.